\documentclass[conference]{IEEEtran}
\usepackage{cite}

\usepackage[utf8x]{inputenc}

\ifCLASSINFOpdf
  \usepackage[pdftex]{graphicx}
  \DeclareGraphicsExtensions{.pdf,.jpeg,.png}
\else
\fi
\usepackage{url}

\begin{document}
%
\title{Repeatable and Reproducible Wireless Networking Experimentation through Trace-based Simulation}



%
\author{\IEEEauthorblockN{Vitor Lamela,
Helder Fontes,
Tiago Oliveira, 
Jose Ruela,
Manuel Ricardo,
Rui Campos}
\IEEEauthorblockA{INESC TEC and Faculdade de Engenharia\\
Universidade do Porto,
Porto, Portugal\\ Emails: \{vitor.h.fernandes, helder.m.fontes, tiago.t.oliveira, jose.ruela, manuel.ricardo, rui.l.campos\}@inesctec.pt}}


\maketitle

\begin{abstract}

To properly validate wireless networking solutions we depend on experimentation. Simulation very often produces less accurate results due to the use of models that are simplifications of the real phenomena they try to model. Networking experimentation may offer limited repeatability and reproducibility. Being influenced by external random phenomena such as noise, interference, and multipath, real experiments are hardly repeatable. In addition, they are difficult to reproduce due to testbed operational constraints and availability. Without repeatability and reproducibility, the validation of the networking solution under evaluation is questionable.

In this paper, we show how the Trace-based Simulation (TS) approach can be used to accurately repeat and reproduce real experiments and, consequently, introduce a paradigm shift when it comes to the evaluation of wireless networking solutions. We present an extensive evaluation of the TS approach using the Fed4FIRE+ w-iLab.2 testbed. The results show that it is possible to repeat and reproduce real experiments using ns-3 trace-based simulations with more accuracy than in pure simulation, with average accuracy gains above 50\%.

\end{abstract}

\begin{IEEEkeywords}
Wireless Networking Experimentation, Trace-Based Simulation, Repeatability and Reproducibility.
\end{IEEEkeywords}

%
\IEEEpeerreviewmaketitle

\section{Introduction}

The increasing need of wireless communications in emerging scenarios, including aerial and maritime, requires the development of new networking solutions. To properly validate them, we depend on performance evaluation of wireless networks, traditionally considering Simulation and Experimentation. Simulations are flexible but usually produce optimistic results, due to many simplifications required to represent complex scenarios such as those related to emerging aerial networks. This lack of accuracy forces experimentation on testbeds to more accurately validate the solution under evaluation. Testbeds, on the other hand, are increasingly costly to maintain and are frequently unavailable; still, they provide realistic results and are an essential step to achieve proper protocol validation and fine tuning. The problem is that real wireless testbed experiments in emerging networking scenarios are hardly repeatable. Given the same input, they can produce very different output results, since wireless communications are influenced by external random phenomena such as noise, interference, and multipath, which result in very unstable radio link quality -- herein represented by the Signal to Noise Ratio (SNR) at the receiver. Real experiments are also difficult to reproduce. Either the original community testbed can be unavailable – offline or running other experiments – or the custom testbed becomes inaccessible. Without repeatability and reproducibility, the validation of the wireless networking solution under evaluation is questionable.

What if we could make any wireless experiment repeatable and reproducible under the same exact conditions? What if we could share the same testbed execution conditions among an "infinite" number of users? What if we could run wireless experiments faster than in real time?

INESC TEC proposed and is developing the Trace-based Simulation (TS) approach \cite{fontes2017trace}\cite{fontes2018improving} that combines the best features of simulation and experimentation. By relying on Network Simulator 3 (ns-3) and its good simulation capabilities from the MAC to the Application layer, we are exploring how ns-3 can be used to replicate real-world wireless experiments. The TS approach introduces new mechanisms to capture the execution conditions of an experiment and enable its repetition and reproduction using ns-3.

The original contribution of this work is the evaluation and validation of the TS approach taking advantage of the high quality and numerous resources provided by the Fed4FIRE+ community testbeds, in the context of the SIMBED project approved in Open Call 3~\cite{fed4fire}. The validation of the TS approach in a larger scale than in \cite{fontes2017trace}\cite{fontes2018improving} will increase the confidence of the networking community in using this approach. Also, it will foster the cooperation between simulation and experimentation communities towards a paradigm shift in the evaluation of wireless networking solutions. The TS approach enables: 1) closer-to-real evaluation conditions, bringing the usefulness of real tesbeds to a broader audience; 2) repeatable and reproducible wireless networking experiments; 3) concurrent user access to the same testbed conditions; 4) offline experiments, even if the testbed is not available or does not exist anymore; 5) faster than real-time experiments; 6) confirming results from scientific publications based on the traces shared by the authors.

The paper is structured as follows. In Section \ref{sec:relatedWork} we present the related work focused on repeatability and reproducibility of experiments. In Section \ref{sec:tsApproach} we describe the TS approach. In Section \ref{sec:trace_validation} we validate the TS approach using the Fed4FIRE+ w-iLab.2 testbed. Finally, in Section \ref{sec:conc} we draw the main conclusions and point out the future work.

\section{Related Work}\label{sec:relatedWork}

To achieve repeatability and reproducibility of experiments, different approaches have been proposed in the literature. 

For experimentation, the CONCRETE \cite{keranidis2012concrete} tool used in Fed4FIRE+ testbeds runs the same experiment multiple times and selects the results representing the system stable operation. In \cite{Papadopoulos_16.5:2016} the authors concluded that although experimentation is more realistic, only 16.5\% of the papers they analyzed had reproducible results. In \cite{Papadopoulos:2016} an approach that runs the same experiment multiple times, spaced in time, and assumes it is valid if the experiment remains reproducible. In \cite{inria2018} the authors proposed a methodology to test whether the results are realistic enough to be considered valid and to allow their reproduction using the same or other testbeds. The problem of all these approaches is that they heavily depend on the testbed stability and availability for multiple runs.

An example of the emulation approach is Mininet-WiFi~\cite{mininet-wifi}. Based on the mininet emulator, Mininet-WiFi supports replaying the position of nodes and the Received Signal Strength Indication (RSSI); Still, only Emulation Mode and symmetrical Wi-Fi links are supported. According to \cite{fontes2017far}, Mininet-WiFi is lacking the support for Minstrel rate control algorithm~\cite{minstrel}, channel contention mechanisms (e.g., CSMA/CA), MAC layer retransmission, and interference.

For simulation, two main approaches can be found: 1) \textit{Packet Based Replay}, such as the one proposed in \cite{PacketBasedReplay}, where the authors capture traffic of real networks and try to reproduce the same experimental condition in simulation down to per packet resolution; 2) \textit{Application Layer Replay}, as the one presented in \cite{TraceReplayAppLayer}, where the authors try to abstract all low level variables and reproduce the traffic delays and performance bottlenecks experienced in the real network at the Application layer. These approaches do not allow to keep improving the solution under evaluation.

To the best of our knowledge, the TS approach originally proposed in~\cite{fontes2017trace}\cite{fontes2018improving} remains the only one that allows to replay the conditions of the scenario both in simulation and emulation mode.

\section{Trace-based Simulation Approach}\label{sec:tsApproach}

In this section we present an overview of the TS approach~\cite{fontes2017trace}\cite{fontes2018improving}, explaining its fundamental aspects.

\begin{figure}
\centering
\includegraphics[width=0.9\linewidth]{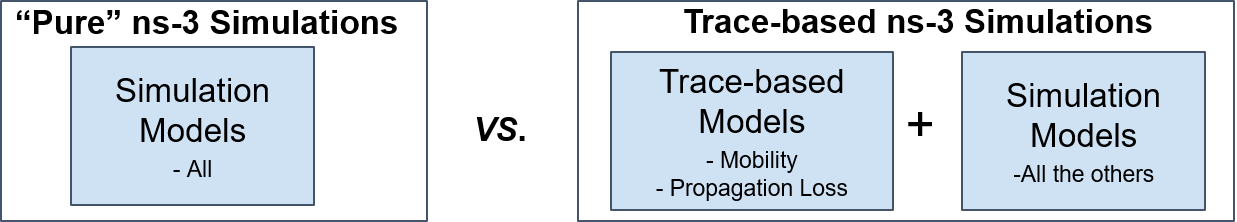}
\caption{High-level comparison between pure simulation and TS approaches.}
\label{fig:pure_vs_trace_comparison}
\end{figure}

The TS approach aims at improving the simulation accuracy as a means to achieve repeatability and reproducibility of past real experiments, thus allowing to continue fine-tuning and evaluating a networking solution through more accurate simulations. The TS approach focuses on capturing traces of the execution conditions of an experiment and enabling the repetition and reproduction of such conditions using the past traces in ns-3. To achieve this goal the TS approach relies on the ns-3 good simulation capabilities from the MAC to the Application layer, combining them with the reproduction of traces that characterize relevant physical parameters, such as the variation of the position of the communications nodes and the quality of the radio links over time, to create realistic simulations. Figure \ref{fig:pure_vs_trace_comparison} depicts a high-level comparison between a pure ns-3 simulation and the TS approach; in the latter, the ns-3 simulation remains the same except for the Mobility and the Propagation Loss Models used. By only replaying and reproducing the experimental conditions that are complex to model in pure simulation, due to their highly unstable and unpredictable nature, the TS approach allows the evaluation of an unlimited number of solutions in the same exact conditions. Below, we discuss how the mobility of nodes and the radio link quality are captured and reproduced using the TS approach.

\textbf{Mobility of Nodes.} In a real-world testbed, the position of the nodes can be frequently changing throughout the experiment. As such, for mobile nodes, this value shall be collected periodically. This is enough to move the nodes in ns-3, according to the real waypoints, as ns-3 natively supports this functionality by using the \textit{WaypointMobilityModel}.

\textbf{Radio Link Quality.} In a real-world testbed, the radio link quality is constantly changing; thus, it must be collected periodically. The SNR at the receiver should be collected at both ends of the radio link, as very often radio links are asymmetric \cite{kurth2006multi}. The SNR is the variable that better represents the radio link quality. The SNR directly correlates to the Bit Error Ratio (BER), which coupled with the frame size and the PHY rate provides the Frame Error Ratio (FER) -- probability of a frame being received with errors and discarded. Simulators also use the SNR to calculate the FER based on error models such as NistErrorRateModel~\cite{pei2010validation}. Based on the FER, ns-3 uses random variable streams to decide whether the frame is successfully received. Frames being dropped result in MAC layer retransmissions, which lowers the throughput and increases delay. Retransmissions can also trigger the auto-rate adaptation mechanism (e.g., Minstrel~\cite{minstrel}) to lower the PHY rate, further reducing the throughput and increasing the delay. Because of the realism of ns-3, we assume that reproducing the same SNR in ns-3 enables more accurate representation of the radio link quality than using theoretical path loss models, achieving FER and link behaviour similar to the real experiment.

\section{Evaluation of the Trace-based Simulation Approach in Fed4FIRE+ w-iLab.2} \label{sec:trace_validation}

The TS approach was evaluated considering a set of experiments over the Fed4FIRE+ w-iLab.2 testbed. These experiments were replayed via their physical condition traces in trace-based ns-3 simulation. Pure ns-3 simulations were also performed to establish a comparison baseline. In the end, the accuracy of the TS approach was compared against pure simulation (PS) approach, considering as reference the experimental results obtained for two network performance metrics (PM): throughput and Round-Trip Time (RTT). The accuracy of the TS and pure simulation approaches was measured using the relative error for the throughput and the absolute error for the RTT metrics. The absolute error is calculated using Equation~\ref{eq:1} and the relative error is calculated using Equation~\ref{eq:2}, where $PM_i$ is the $PM$ value obtained for the approach $i$ (TS or PS) and $PM_e$ is the $PM$ value obtained in the real experiment. The accuracy gain of the TS approach with respect to the PS approach is calculated using Equation~\ref{eq:3}, where $RelativeError_{TS}$ and $RelativeError_{PS}$ are the relative errors for TS and PS approaches, respectively.

\begin{equation} \label{eq:1}
Absolute  Error_i = |PM_i - PM_e|
\end{equation}

\begin{equation} \label{eq:2}
Relative  Error_i = \frac{Absolute  Error_i}{PM_e} \times 100 (\%)
\end{equation}

\begin{equation} \label{eq:3}
AccuracyGain = \bigg(1 - \frac{RelativeError_{TS}}{RelativeError_{PS}}\bigg) \times 100 (\%)
\end{equation}

\subsection{Experimental Setup}

To extensively evaluate the TS approach we ran a large number of Wi-Fi experiments. The experiments considered Wi-Fi point-to-point links established between a static transmitter and a static receiver, using auto PHY rate mode. Because the SNR is the metric that better represents the impact of the testbed physical conditions on the radio link quality -- and the resulting throughput and RTT -- we ran experiments to measure the network performance from very low to very high SNR conditions. This allowed testing the TS approach in a broad performance range of a Wi-Fi point-to-point link.

For the experiments we reserved and used a selection of Zotac Wi-Fi nodes from the w-iLab.2\footnote{https://inventory.wilab2.ilabt.iminds.be/?viewMode=inventory [Accessed: 28th January 2019]} testbed. The Zotac nodes are placed in a grid pattern, with a column width of 6~m and row height of 3.6~m. These nodes have an Intel Atom D525 CPU (2 cores, 1.8 GHz), 4 GB of RAM and 2 IEEE 802.11abgn Sparklan Wi-Fi interfaces with AR9280 Atheros chipset. Attached to the AR9280 interfaces the nodes have 3~dBi gain dipole antennas with 10~dB attenuators inline -- adding a total of 20 dB to the path loss -- to avoid saturating the receivers and limit the testbed interference. The TX power of these interfaces can be controlled from 0 to 17 dBm; this was very important to use the same set of nodes to create 18 different experiments when it comes to the SNR levels at the receiver. The Ubuntu 14.04 LTS x64 OS was used.

For each experiment we collected the following data per node: 1) traces of real SNR at the receiver for each received frame, organized by peer node and time-referenced with a microsecond resolution; 2) the position of the nodes, once per experiment as the nodes are static; 3) network PMs measured for each link -- average throughput and RTT. The traces of real SNR and node positions are used to feed the trace-based ns-3 simulations. The network PMs are used to evaluate the TS approach accuracy against the PS approach. 

The TS approach focuses on reproducing the experimental physical conditions. However, the network performance can also be influenced by the number of queues and respective sizes, and the related traffic control and queue management mechanisms used in the nodes. If the conditions in the real experiments and ns-3 simulations are different we may end up with rather different results. For this reason, we focused on measuring the PMs in the following conditions:

\textbf{Throughput.} Measured at the receiver, considering the sender is generating traffic with offered load that saturates the link. This assures that there are always packets queued waiting to be sent.

\textbf{RTT.} Measured using ICMP echo requests and replies, without concurrent network traffic, so that the queues are empty when a packet is generated. In this way we are only considering the delays related to the access and (re)transmissions over the Wi-Fi half-duplex multiple access wireless medium, and discarding any queuing and processing delays.

To carry out the aforementioned experiments over w-iLab.2 we used the following methodology:  

\textbf{1) Reservation of Nodes.} We started by reserving 4 consecutive Zotac nodes in the same grid row. The leftmost node was the \textit{Master}; \textit{ClientA}, \textit{ClientB} and \textit{ClientC} nodes were placed at 6, 12 and 18 m, respectively, from the \textit{Master}.

 \textbf{2) Startup of Nodes.} We selected our custom Ubuntu 14.04 LTS x64 OS image to boot in all nodes, containing our experimentation scripts and patched ath9k driver.

\textbf{3) Configuration of Nodes.} We made the following configurations: \textit{NTP client} configured so that all the nodes were clock synchronized; \textit{Wi-Fi standard} set to IEEE 802.11a; \textit{Wi-Fi operation mode} set to ad-hoc; \textit{Channel bandwidth} set to 20 MHz; \textit{Channel center frequency} set to 5220 MHz, which remained free during all the experiments; \textit{PHY rate} set to auto PHY rate mode; \textit{TX power} set from 0 to 17 dBm in 1 dBm steps, according to the experiment.

\textbf{4) Run batches of experiments.} We started by configuring all the 4 nodes with the same TX power. For each TX power value we ran 300 s experiments, considering the communication between the \textit{Master} node and a single \textit{Client}, one at a time. We tested the following three scenarios: \textbf{a) Idle network link between the \textit{Master} and the \textit{Client}} using the \textit{ping} application to measure the RTT, generating 10 requests per second with a packet size of 1472 bytes; \textbf{b) Unidirectional UDP flow} using the \textit{iperf3} application to sequentially generate an UDP flow from \textit{Client} to \textit{Master}, and then from \textit{Master} to \textit{Client} to test link asymmetry, with offered load (54 Mbit/s) above link capacity (28-30 Mbit/s); \textbf{c) Bidirectional UDP flows between \textit{Master} and \textit{Client}} generating two concurrent UDP flows in opposite directions.

\subsection{Trace-based and Pure Simulation Results}

To evaluate the TS approach accuracy, all the real experiments were reproduced in ns-3 feeding the \textit{TraceBasedPropagationLossModel} with the real traces of SNR. The same network PMs were measured in ns-3 (throughput and RTT). To have a baseline of comparison, we reran the equivalent simulations using the PS approach, where we tested four different path loss models:

\textbf{1) \textit{FriisPropagationLossModel}}\cite{FRIIS}\textbf{.} As the nodes have radio line-of-sight, and the direct radio ray is expected to be dominant due to the low distance and the absence of obstacles in between the nodes. This path loss model is deterministic, and thus the SNR remains constant along the simulation.

\textbf{2) \textit{LogDistancePropagationLossModel}}\cite{logDistance}\textbf{ ($\gamma$ = 2.0) plus Rician fast fading}\cite{rician}\textbf{.} With $\gamma$ = 2.0 the \textit{LogDistance} model has the same output as the \textit{Friis} model; this model adds the Rician fast fading by configuring the \textit{NakagamiPropagationLossModel}~\cite{NAKAGAMI}  with m = 1.25. 

\textbf{3) \textit{LogDistancePropagationLossModel} ($\gamma$ = 1.7) plus Rician fast fading:} As the w.iLab.2 is an indoor testbed, has radio line-of-sight, and may have a strong multipath component that adds substantially to the direct ray, we decided to test a reduced path loss exponent $\gamma$. 

\textbf{4) \textit{LogDistancePropagationLossModel} ($\gamma$ = 2.5) plus Rician fast fading.} To complement the two previous path loss model options, we considered a higher path loss exponent $\gamma$. 

Apart from the ns-3 \textit{PropagationLossModel} used, all other simulation parameters are the same for both simulation approaches, except for the ``RF gain'' which is set to 0~dB in the case of the TS and set to -7~dB for the PS. This is to adjust the path loss calculation considering the 3 dBi gain of the antennas and the 10 dB attenuation of each inline attenuator ($3 - 10 = -7$ dB). Note that the ``RF gain'' is considered on both ends of the communication, so the resulting ``RF gain'' becomes -14 dB, which is added to the path loss calculation. In ns-3, we generated the UDP traffic using the ns-3 \textit{OnOffApplication} traffic generator and the RTT measurements were performed using the ns-3 \textit{V4Ping} tool.

For each real second of a given experiment, and their corresponding TS and PS counterparts, we compared: 1) the average throughput per second (kbit/s); 2) the median of the RTT samples per second (ms). By comparing these two network PMs for the exact same time interval considering the real experiment, TS, and PS we calculated the relative error for the TS and PS approaches using Equation~\ref{eq:2}. We found this method to be the most adequate to calculate the relative error, as we are trying to reproduce real Wi-Fi experiments with performance variations along the time.     

\begin{figure}
\centering
\includegraphics[width=1.0\linewidth]{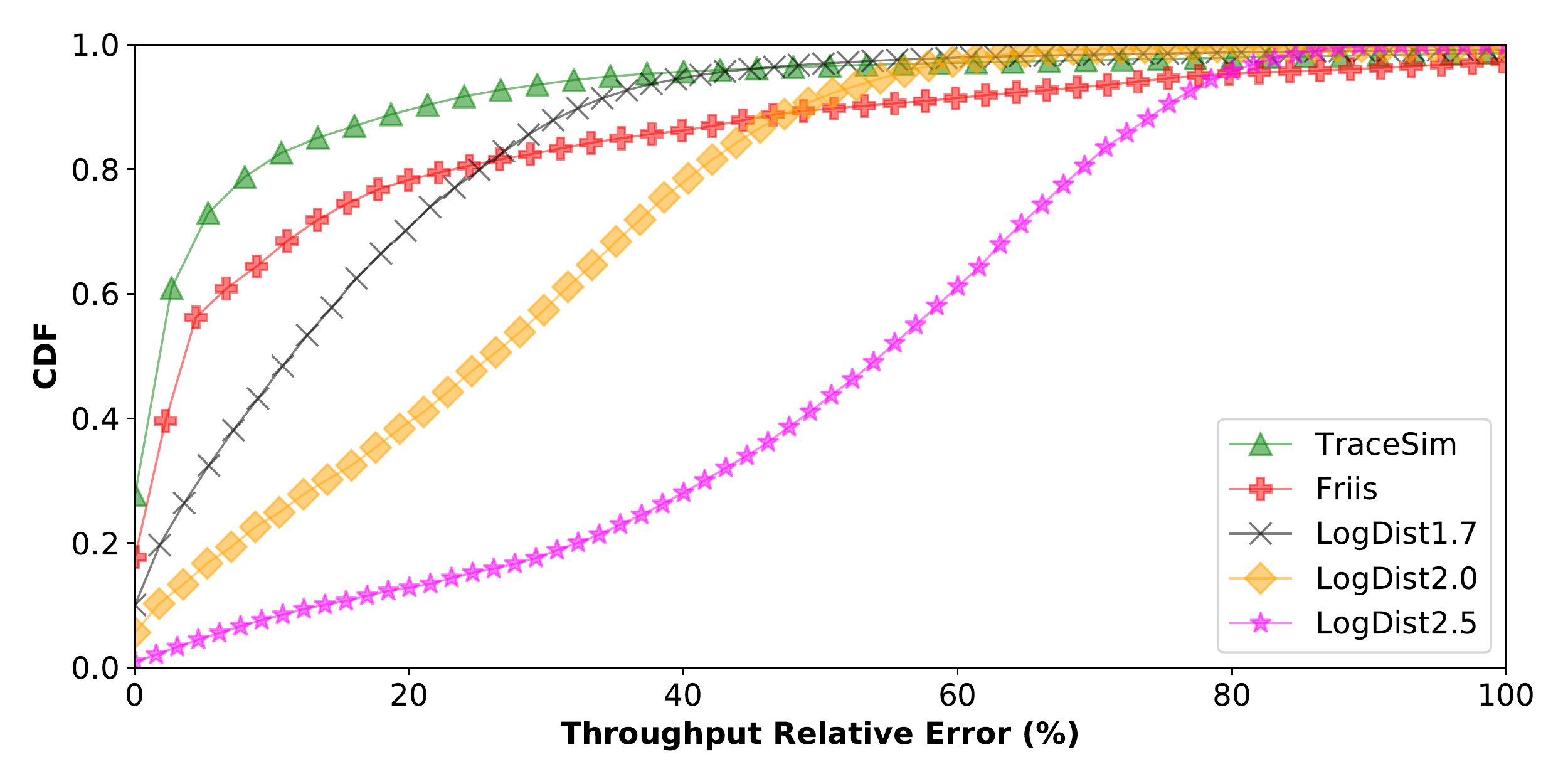}
\caption{CDFs of the throughput relative error when comparing the trace-based (TraceSim) and pure simulations to the corresponding real experiments.}
\label{fig:autoGoodputError}
\end{figure}

Figure \ref{fig:autoGoodputError} shows the CDF of the throughput relative error for the TS and PS when the Wi-Fi point-to-point link is running in auto PHY rate mode. For computing the CDFs, all samples with real throughput equal to 0~kbit/s were discarded to filter the initial seconds in the experiment where \textit{iperf3} did not yet start sending traffic. In a total of 31058 samples, the filtered samples represent 1.2\%. Table~\ref{tab:autoGoodputError} summarizes the relevant values extracted from the CDF plot and presents the 90th percentile, 50th percentile (median), and the average throughput relative error. Analyzing the pure simulation results we can observe that between the four options the Friis model and the LogDistance model with $\gamma$~=~1.7 plus Rician fast fading (LogDist1.7) are the ones that better approximate the real experiment results. This shows that the Friis path loss model, although not considering fast fading, is the one that, on average, more closely matches the real experiment results. This was expected considering the isolated and very stable w-iLab.2 testbed. Analyzing the TS results, we can observe that it is the one that more closely reproduces the real experiment: for the 90th percentile, the TS approach presents an accuracy gain (c.f. Equation \ref{eq:3}) of 70\% over Friis and 56\% over LogDist1.7; for the median, the TS approach has an accuracy gain of 17\% over Friis and 62\% over LogDist1.7; finally, on average, using the TS approach there is an accuracy gain of 56\% over Friis and LogDist1.7. In conclusion, the results show that the TS approach is considerably better at reproducing a closer-to-real throughput than a PS approach using a typical path loss model.

\begin{table}[h]
\centering
\caption{Throughput relative error when comparing the trace-based (TraceSim) and pure simulations to the corresponding real experiments.}
\label{tab:autoGoodputError}
\begin{tabular}{c|c|c|c|}
\cline{2-4}
\multicolumn{1}{l|}{} & \multicolumn{3}{l|}{\textit{\textbf{Throughput Relative Error (Auto PHY Rate) (\%)}}} \\ \cline{2-4} 
\multicolumn{1}{l|}{} & \textbf{90th Perc.} & \textbf{50th Perc. (Median)} & \textbf{Average} \\ \hline
\multicolumn{1}{|c|}{\textbf{TraceSim}} & 14 & 5 & \textbf{7} \\ \hline
\multicolumn{1}{|c|}{\textbf{Friis}} & 46 & 6 & 16 \\ \hline
\multicolumn{1}{|c|}{\textbf{LogDist2.0}} & 50 & 31 & 29 \\ \hline
\multicolumn{1}{|c|}{\textbf{LogDist1.7}} & 32 & 13 & 16 \\ \hline
\multicolumn{1}{|l|}{\textbf{LogDist2.5}} & 77 & 58 & 54 \\ \hline
\end{tabular}
\end{table}

\begin{figure}
\centering
\includegraphics[width=1.0\linewidth]{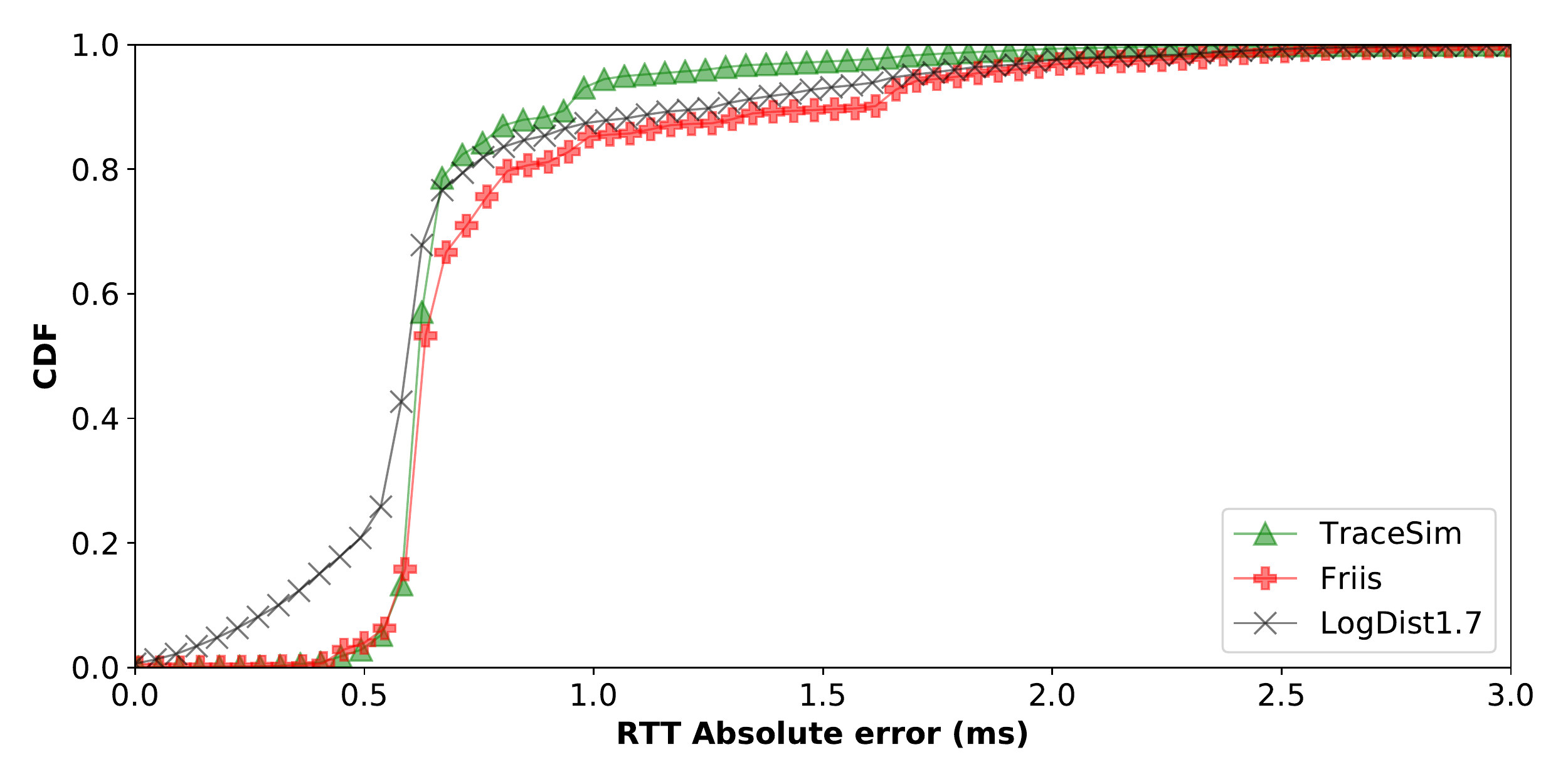}
\caption{CDF of the trace-based ns-3 simulation (TraceSim) and pure ns-3 simulation (PureSim) RTT absolute error in comparison to the RTT obtained in the real experiments with packet size of 1472 bytes.}
\label{fig:rttDelayAbserror}
\end{figure}

Regarding RTT, we chose to represent the absolute error instead of the relative error (cf. Equation~\ref{eq:1}), as very small differences between low RTT values would give very high relative errors. For the PS approach, only the Friis and LogDist1.7 options were considered as they were the most accurate for the throughput. Figure \ref{fig:rttDelayAbserror} shows the CDF of the absolute error of the RTT measured for TS and PS in comparison to the RTT obtained in each second of the real experiments with packet size of 1472~bytes. In the case of the RTT we considered the median value for each second, excluding the seconds in which we could not get a delay sample in either the real experiment, TS, or PS, as we can only compare samples containing RTT measurements. The filtered results account for roughly 1.5\% of all the one second samples. In the plot we can observe that the median of the error for all simulated scenarios is around 0.6~ms, even for the TS approach. In this case, the TS approach suffers from the same problem of the pure simulation: the fact that ns-3 does not account for the nodes processing time in the protocol stack, contrary to real nodes where the RTT is the sum of processing time and transmission time. This is confirmed by the minimum value of RTT measured in the real experiments and in ns-3, which were around 1.1~ms and 0.5~ms, respectively. From this we can infer that the processing time is around 0.6~ms, explaining the median absolute error. Taking this aspect into consideration, we can conclude that having RTT absolute error values represented in the CDF curve below the 0.6~ms error mark is as indication that the transmission time was higher than in the real experiment, due to more frame retransmissions or the use of a lower PHY rate than in reality. Based on this, we can see that, once again, the TS approach is better than the PS approach alternatives. For the 90th percentile the TS approach lowered the absolute error from 2.3 and 1.9~ms, respectively from Friis and LogDist1.7, to 1.5~ms. After analyzing the results we can conclude that the TS approach provides RTT values closer to the real experiments than a PS approach. This accuracy gain over the PS approach is due to the fact that it reproduces a more accurate SNR, which enables more realistic frame retransmissions and the usage of closer-to-real PHY rates.

Overall, we can conclude that even for the w-iLab.2 scenarios -- where we are dealing with static nodes -- the TS approach brings significant gains over the PS approach, successfully reproducing closer-to-real network performance results by considering the SNR observed in the real experiment. By using the TS approach in emerging testbed scenarios we expect to get even higher gains when compared to the PS approach, considering their highly complex and unstable nature.

\section{Conclusions and Future Work}\label{sec:conc}

The TS approach was extensively evaluated using a large set of experiments over the Fed4FIRE+ w-iLab.2 testbed. The evaluation results showed that the TS approach has significant accuracy gains when compared to the use of a PS approach; for the throughput PM it achieved average and 90th percentile accuracy gains above 56\%. This is especially relevant, considering the fairly controlled and static scenario of w-iLab.2. We expect the TS approach to present even higher accuracy gains for emerging networking scenarios, where PS models are less accurate. The TS approach enables: 1) concurrent user access to the real testbed conditions based on past traces; 2) running simulations faster than in real time; 3) running multiple simulation instances at the same time, exploring different variants of the solution under evaluation. Using the TS approach, it is also possible to reproduce the same experiment in real-time, connected to external real nodes, which allows to keep improving and fine-tuning client systems that depend on the communications system to operate. These advantages have the potential to foster the interaction between simulation and experimentation communities, with mutual benefits. 

As future work, we are currently working on a framework to assist the
related processes of capturing, reusing, and sharing traces. We are also working on a Machine Learning (ML) approach to fine-tune current simulation models and create new models based on the real traces. With this ML approach we aim at bringing the accuracy gains of the TS approach to PS scenarios that are more flexible (i.e, enable different scale and mobility) than only replaying the past experiments.

\section*{Acknowledgments}

This work is financed by national funds through the FCT - Foundation for Science and Technology, I.P., under the project: UID/EEA/50014/2019. The second author would like to thank the support from the FCT under the fellowship SFRH/BD/69051/2010. This work is also part of the SIMBED project.



%

\bibliographystyle{IEEEtran}
\bibliography{IEEEabrv,EuCNC2019}

\end{document}